# Vulnerability analysis of three remote voting methods

Chantal Enguehard & Rémi Lehn
Université de Nantes
Laboratoire d'Informatique Nantes Atlantique
2, rue de la Houssinière
BP 92208
44322 Nantes Cedex 03
France

with the support of the European Computer and Communication Security Institute
Bruxelles, Belgique

**Resume**
This article analyses three methods of remote voting in an uncontrolled environment: postal voting, internet voting and hybrid voting. It breaks down the voting process into different stages and compares their vulnerabilities considering criteria that must be respected in any democratic vote: confidentiality, anonymity, transparency, vote unicity and authenticity. Whether for safety or reliability, each vulnerability is quantified by three parameters: size, visibility and difficulty to achieve. The study concludes that the automatisation of treatments combined with the dematerialisation of the objects used during an election tends to substitute visible vulnerabilities of a lesser magnitude by invisible and widespread vulnerabilities.

**Key-words :** Internet voting, remote voting, postal remote voting, hybrid remote voting, democracy, transparency, fraud, anonymity, authenticity, unicity, visibility, virus, worms.

## Introduction

Remote voting procedures have been renewed recently with the introduction of optical scanners to automatically read the ballots or to completely dematerialise the objects used to vote by an internet voting process. This article studies three methods of remote voting (postal voting, hybrid voting and Internet voting). It describes the various phases. Technical vulnerabilities of internet voting are set out in part three, while the fourth part compares the vulnerabilities of each type of vote.

## I. Remote voting

### I.1 - Definition

Depending on the country, remote voting may consist of two separate concepts:

— Voting is supervised but takes place outside the normal location (e.g in an embassy);
— Voting takes place in an uncontrolled environment and in the absence of any electoral officer.

We are interested here in remote voting outside the control of an electoral officer in the following three forms: Internet voting, postal voting and hybrid voting.

The scope of a study of the elections may include the preparation of voter lists, the candidates' campaign up until the announcement of results. We focus here only on the ballots that we observe from their delivery to the voters until the counting of votes.

We do not present questions relating to paper voting procedure that have already been studied (see [7] and [15]), or aspects of the digital divide and accessibility (see [3], [14]).

## I.2 - Three ways to vote remotely in an uncontrolled environment

For each mode of remote voting, we define a model represented by a real application widely used and which we consider as representative of the practices.
— Internet voting: Internet voting procedure used in the canton of Geneva in 2007 [10].
— Postal voting: as used in the canton of Geneva in 2007 [31].
— Hybrid voting: hybrid voting procedure used in the elections of the Comité National de la Recherche Scientifique (CNRS) in France in 2008.

### Internet voting

Internet voting (i-voting) is part of a broader package called electronic voting (e-voting). Under the latter are grouped all forms of voting involving an electronic device to cast or count votes.

There are drafts of standards and international norms but they lack precision in their definition of the necessary organizational, legal and technological models. There are, therefore, many different Internet voting procedures. However, it is possible to expose a general pattern that is more or less respected by the usual procedures of Internet voting that are said to be secure. Information relevant to authentication are provided to voters by mail. Voters log on an official web site to vote from any computer connected to the Internet and equipped with a browser compatible with the application running on the official web site. Each voter uses the information that she had previously received to be identified (login and password), and then she express her choice. It is encrypted and sent to the server hosting the official web site that collects the votes, stores them until the close of the poll and produces the results of the vote at the close of the poll.

Because all the voters do not have a computer with an Internet connection, this method of voting is always an addition to a postal voting procedure[1].

### Postal voting

Each voter receives the material for voting by mail. It includes a "voting card"[2] bearing the identity of the voter, the correspondence envelope and an anonymous envelope. To vote, the voter puts the ballot of his choice in the anonymous envelope that she seals. Then she slips this envelope and the voting card that she dates and signs into the correspondence envelope. This correspondence envelope is then sent by post to the election office.

The election office collects the envelopes as they are received. The counting takes place in two phases. First, the names of the voters are ticked off on the signature register. Then, the correspondence envelopes are opened to collect the anonymous envelopes that are randomised to break any link between them and the envelopes of correspondence. Finally they are opened, ballots are extracted and votes are counted in order to determine the outcome of the vote.

---

1  With the notable exception of France where the decree n° 2007-554 of the 13th of April 2007 on detailed rules for the electronic election on the order of nurses precises that "Electronic voting precludes any other method of voting."
2  The term "voting card" is polysemic. Here it is a paper card bearing the name and the address of the voter.

**Hybrid voting**

The hybrid voting procedure is a modification of the postal voting procedure to allow for automated counting. Voters receive electoral materials by mail: a "voting card" and a single envelope. Each card carries a voting mark (barcode and/or number) to identify the voter and a series of boxes placed in front of the proposed alternatives. The voter blackens the boxes of her choice to vote.

The election office collects the envelopes as they are received. The counting is automated: voting cards are extracted from the envelopes and then scanned. A computer updates the signatures registry to be marked and the number of votes obtained by each candidate.

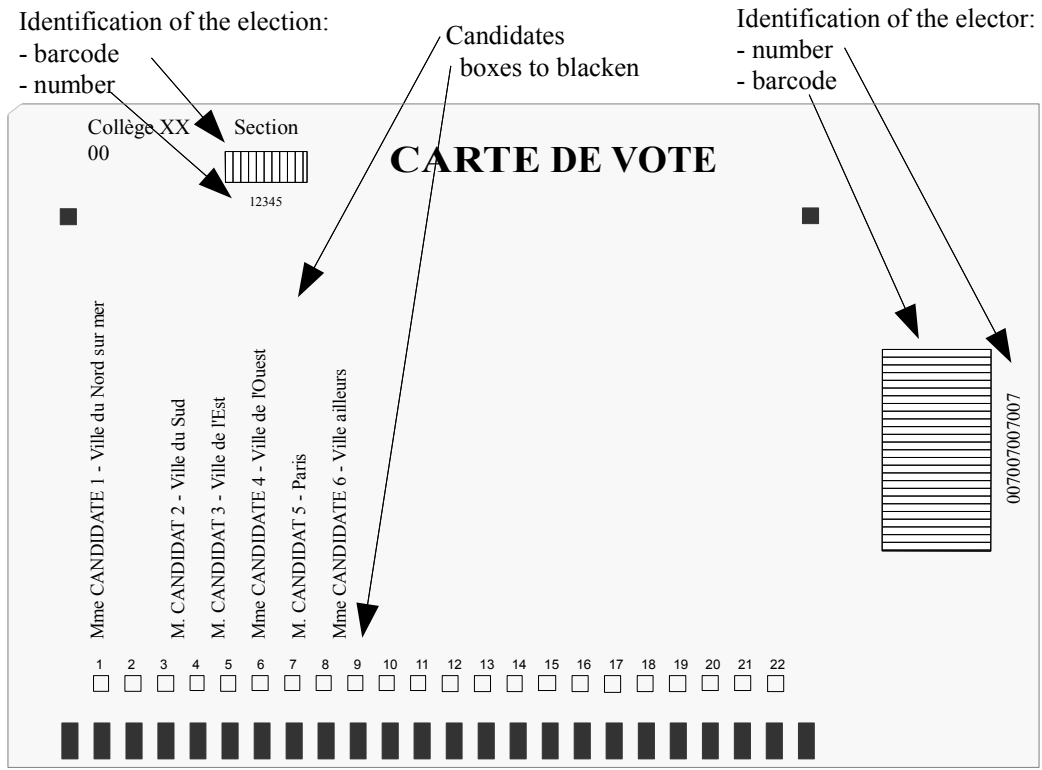

*figure 1 : Voting card for hybrid voting*

## I.3 - Phases of remote voting in an uncontrolled environment

Remote voting in an uncontrolled environment follows a path that can be split into several abstracted phases that are common to the three methods of voting we are observing, but implemented differently depending on the voting method (see table 1): the organizers of the vote prepare electoral material (B1), and its transmission (B2). The electoral material travels through the transmission channel (C1) and is received by the voter (E1). Voters express their choice (E2) and then prepare to send their vote (E3). The ballot is transmitted (C2). The polling station receives ballots (B3) and then performs the necessary counts (B4).

This presentation does not include all communications, for example, Internet voting involves several communications between the voter and the voting system during the vote decision phase (E2).

|  |  | **Internet voting** | **Postal voting** | **Hybride voting** |
|---|---|---|---|---|
| B1* | Preparation of electoral material | Drawing up lists of identifiers and passwords | Printing of electoral material | Printing of electoral material |
| B2* | Preparation for the dispatch of electoral material | Fold up and transfer to the post office | | |
| C1* | Transmission of electoral material | Login and password are sent by post | The two envelopes and the ballots are sent by post | The envelope and the voting card are sent by post |
| E1 | Receipt of electoral material | The electoral material is received by the elector | | |
| E2 | Expression of choice | The voter connects to the electoral web site, registers, authenticates, makes her choice and confirms | The voter express her choice through the ballot paper | The voter express her choice through the voting card |
| E3 | Preparation for sending | The virtual ballot is encrypted | The voter puts her ballot paper in the anonymous envelope, and then in the correspondence envelope | The voter puts her voting card in the enveloppe |
| C2 | Transmission of the ballot | The ballot travels to the officiel server through Internet | The post transport the enveloppes containing ballots to the polling office | |
| B3 | Reception | The official web site stores the received envelopes, updates the signatures list and return receipts to voters | The polling station receives and stores the envelopes containing ballots | |
| B4 | Counts | The software decrypts and count the votes | The signature register is updated, the anonymous envelopes are opened and the votes are counted | The envelopes are opened, the scanner reads the ballot, the software updates the signature register and counts the votes |

\* This step can be non-existent when voters connect with a connection card with a magnetic stripe, as in Estonia.

*figure 2 : Phases of remote voting in an uncontrolled environment*

## II. Methodological choices

### II.1 - Comparative Approach

All voting systems have vulnerabilities, there is no perfect voting system that ensures strict compliance with the principles of democratic elections and gives entirely fair results. Our analysis will compare three models of remote voting according to criteria expressed by various international organizations: the Universal Declaration of Human Rights (Article 21) [23], the Code of Good Pratice in Electoral Matters of the Venice Commission [8], the Election Observation Handbook of the Organization for Security and Cooperation in Europe (OSCE) [27]. These criteria may be characteristic of any democratic vote or be specific to the remote vote [9].

We quantify the consequences of major weaknesses through three parameters: vote magnitude, visibility and difficulty.

— The vote magnitude depends on the number of votes potentially affected by a fraud or a malfunction. This parameter may be small (a few votes), average (number of votes sufficient to change the outcome of elections) or large (potentially nearly all votes).

— The difficulty is a fuzzy estimation of the likelihood of the occurrence of conditions required to exploit a vulnerability. In the case of a problem of technical reliability, it estimates if the failure is common or rare. For a fraud, it measures the complexity of its successful implementation (number of people involved, technical knowledge, cost, discretion, etc.).. This parameter can take three values ; small, medium and large.

— The visibility is used to measure whether the consequences of a vulnerability are evident or not. It can take three values: zero (consequences are invisible), medium (consequences are visible but can not be proved to a court) or large (consequences are sufficiently visible to render the election null and void).

The worst case scenario occurs when the vote magnitude is large, visibility is zero and the difficulty small. These three criteria are not independent: difficulty and visibility will be estimated for a large or medium magnitude.

## II.2 – Democratic remote voting

### Democratic Elections

Remote voting is part of the governance processes on which democracies are based on. The criteria set out by international agencies seek compliance with the essential qualities of democratic elections:

— Unicity: the 'one elector, one vote' principle[3];

— Confidentiality: each voter expresses her choice alone;

— Anonymity: it is impossible to link a ballot to the voter who cast it[4];

— Sincerity: the results of the election reflect faithfully the will of the voters;

— Transparency: "the system's transparency must be guaranteed in the sense that it must be possible to check that it is functioning properly." (Venice Commission) [9].

### Remote voting

These generic criteria are supplemented by specific criteria to remote voting:

— Safety: the system can withstand prospective attacks;

— Reliability: the system works, in spite of hardware or software deficiencies.

The main difficulty is to ensure that votes are not distorted or lost between the casting of the votes by the voters and the counting of the ballots.

## III - Technical vulnerabilities of Internet voting

Voting by Internet is a new procedure characterized by the dematerialisation of all objects relating to the voting procedure (ballots, ballot box, signing sheet). We describe some technical flaws that may change and distort the virtual entities which represent these objects. These vulnerabilities can concern safety or reliability.

---

3  It is the uniqueness that makes an election being universal. Every person of voting age (and not deprived of his civil rights) can vote once. There are no other criteria limiting the right to vote as it was in France with the "censitaire" (a minimum income was required) or the denial of voting rights to women, still current in some countries.

4  Confidentiality and anonymity are two aspects of the secrecy of the vote.

### III.1 - Safety

**Worms[5] and viruses**

The computer used by the voter is likely to host worms and viruses that can trigger attacks to modify the choice expressed by the voter. Most antivirus softwares can only detect worms and viruses that are already known, new viruses can not be identified before proceeding. In addition, the attackers have the advantage of being able to test their creations using the same commonly distributed anti-virus software that is used by their potential victims. The latest viruses are able to pass firewalls and other defences, and are difficult to detect [22] [30]. Attackers can create new viruses, or viruses that modify existing ones (kits exist on the internet for the construction of viruses). A virus can easily infect a large number of computers without being detected and remain dormant until voting day. Viruses could carry out many undesirable operations, unbeknownst to the voter, such as capturing the server connection details, changing the vote of the elector before encryption, spying on the vote of the electors and disclosing all these details to a third party.

**Pharming**

The voter is a victim of misuse of session when she typed in the address URL[6] the official web site address site and she navigates using the protocol for securing communications SSL[7]. She believed she votes on the official web site when in fact she is interacting with a web site that imitates the official web site including by sending a confirmation of receipt of the vote.

The theft can be unmasked if the voter verifies that the security certificate is known and valid. But a falsified safety certificate may have been accepted on the same computer during a previous connection to a web site thought to be secure, causing the display of a warning (see Figure 3). In this case, many users choose to continue, without being aware that they allow a potentially falsified safety certificate to join safety certificates that have been duly approved by certification authorities. Thus, when connecting to the fake voting web site, there will be no security alert.

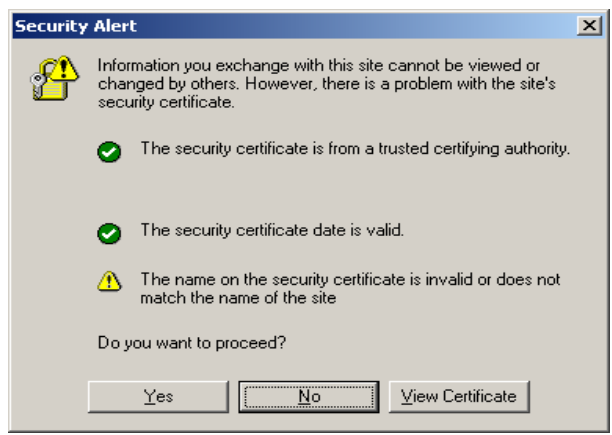

*figure 3 : Window Security Alert*

**Man-in-the-middle**

A man-in-the-middle attack consists of impersonating the server from the point of view of the voter's computer, and impersonating the voter's computer from the point of view of the server. The fraudster can change the cast vote. Encryption voting offers good protection against this attack if the public key encryption that has been sent to the voter has not been intercepted by the fraudster. It is thus necessary to send this key via a secure mail. However it is not necessary to know the encryption key to capture and destroy ballots and to return confirmation messages to the voters to make them believe that their votes had been registered. Voters are deprived from exercising their right to vote without knowing it unless they check the the signatures registry.

---

5   A worm is a virus that has the ability to spread alone by using the network.
6   URL : Uniform Ressource Locator.
7   SSL : Secured Socket Layer.

**Denial of service**

Denial of service is when a vote server is bombarded with connections to prevent legitimate voters from voting. The server, saturated with requests, can not respond to all the demands for connections and is likely to crash.

## III.2 - Reliability

### 1a - Hardware errors

A computer may experience failures or malfunctions. There may be defects in equipment, including electronic cards (faulty welds), or even in microprocessors. Computers must therefore incorporate mechanisms for error detection, which is not routinely done on personal computers.

### 1b - Software errors

There may be errors in programs that run on a computer. These errors can occur at all levels: operating system, softwares, compilers, security vulnerabilities, etc. That is why the National Institute of Standards and Technology (NIST) recommends that the control of the results of an electronic voting system should not be processed by a software application that, too, may experience malfunctions [24]. These results were confirmed by numerous academic studies on dematerialised voting [12] [18] [21] [33], the Irish independent commission on electronic voting [4] [5] or international institutions [26] .

Several ways were explored to detect and eliminate errors in Internet voting: testing, development of formal programs, expertise, monitoring of elections and verification of cryptographic results.

*Tests*

Successfully testing an application can not predict with certainty the behaviour of the same application in future uses, or even its performance during past operation. It is not possible to simulate or reproduce the course of a real election involving thousands of people, with all the hazards that may occur. The process of testing is inadequate to prove the correctness of a computer program. It is, therefore, not suitable for an electronic voting application where malfunctions may go unnoticed because of the anonymity inherent in the system.

*Formal development*

In science terms, to be certain that a program has no errors, it should at least be required to use formal development methods. These methods are very expensive and limited to software components. Beyond a certain level of complexity, there are still no sure development methods[8].

*Control expert*

There may be certification authorities but they lack the ability to verify programs with sufficient resources and attention to detect all errors and security vulnerabilities. Finally, even if such an examination was made, even if we had development methods to avoid human error, there would remain an unsolved problem, namely to ensure that the programs in use are exactly those have been certified or to ensure that these programs run without modification forced by the environment (execution modified by a malicious piece of code introduced by some peripheral software or device). In all cases, the server uses an operating system, possibly a compiler or a code interpreter, that should also be considered, etc. This approach quickly becomes daunting and therefore impraticable[9].

---

8  « We don't have a theory that can guarantee system reliability, that can tell us how to build systems that are correct by construction. We only have some recipes about how to write good programs and how to design good hardware. We're learning by a trial-and-error-process » J. Sifakis [32].

*Monitoring of elections*

To trace the operation of a computer application, its progress must be observed step by step. But, introducing probes into software programs to monitor their performance raises the issue of objectivity and neutrality of these probes and of the programs that analyse observed data. As part of a voting system, such monitoring involves keeping a logbook in which all events are recorded and time-stamped: arrival of a ballot, signing the signing registry, counting, etc. The problem is that reading this logbook would allow everyone's vote to be determined, which constitutes a violation of the voting secrecy. If the information in the logbook is not complete (to protect the secrecy of the vote), the process becomes useless as it is no longer possible to fully monitor the processing of information received and to detect malfunctioning (or fraud). We note here that an effective measure in the context of the usual uses of the Internet (such as bank transactions) can not be successfully implemented because of the very special features of anonymous democratic elections.

*Verification of the results a posteriori*

Voting by Internet is the subject of intense research in the field of cryptography to provide models allowing any voter to verify that his vote is taken into account and that the total of all votes is correct. The elector must also be able to provide evidence for its findings. Some experimental systems have been implemented as RIES [16] or VoteBox [29]. These systems exhibit a high degree of complexity, which is a factor of vulnerability: a well-designed cryptographic protocol may make errors of settlements and be vulnerable to fraud [17], [28]. Moreover, with these systems, even if a voter sees is a distortion of her vote she can not prove it. In addition, to respect the confidentiality it is indispensable to destroy the intermediate files[10].

## IV. Evaluation

Different approaches are possible to structure this evaluation because the analysis must take into account several dimensions: compliance with the criteria that characterize a democratic election, technical characteristics of each mode of voting, or the spatio-temporal sequence of attacks on an election. We will follow this last thread by first addressing issues common to the three methods of voting and then treating them individually.

## IV.1 Common issues

**Preparation, dispatch and transmission of the electoral material**

An incident or a misconduct can lead to a failure in printing (B1) or transmitting (B2) the electoral material to a portion of the electorate. Postal mail containing electoral material can be lost, delayed or diverted during transmission (C1). The voters are thus deprived of their right to vote.

---

9  « les experts ne contrôlent que ce qu'ils veulent, ou ce qu'ils peuvent. » (Experts only control what they want, or what they are capable to) A. Auer [2]

10  It is difficult to destroy files in order to make impossible their reconstruction.

These interferences with the principles of unicity and authenticity may be of average magnitude, its difficulty is small for a person involved in the organization. It presents medium visibility because the letters were not sent by registered post for reasons of cost, there is no control of their issue and inattentive voters are unlikely to notice this non-delivery and to report it officially. A strict control of the number of letters actually sent is essential.

**Receipt of electoral material (E1)**

When a postal mail is received, it can be intercepted by one of the many people sharing the same home. This fraud may be committed by a person close to the voter. This person is likely to know the additional information required to be allowed to vote (usually the date of birth). This fraud has a small magnitude: a fraudster may only divert a few votes.

The biometric processes are often considered to prevent identification fraud for Internet voting. This approach meets different obstacles. First, it contradicts several security principles such as the fact that a password should always be stored in a single file and encrypted, could be changed if necessary and that the stages of identification and authentication should be separate. When biometric procedures are implemented, we observe that the same data is used to identify and authenticate. This data is not secret and it is impossible to change. In addition, it has been repeatedly demonstrated that it is easy to fool biometric systems [20][11]. Finally, generalizing this approach involves identifying and centralizing the biometric data of all voters, which poses technical, organizational and ethical problems.

**Non receipt of electoral material (E1)**

The envelopes carrying the electoral material that did not reach their addressee are returned to the sender, ie the polling station. It may be tempting to use them to vote.
The magnitude of this fraud is limited by the number of envelopes returned to the polling station. It is important that these envelopes are counted and that number is noted in the official minutes as a measure to identify a large scale diversion of electoral material.

**Expression of choice (E2)**

Respecting confidentiality means that voters vote alone and without any coercion. None of the remote voting modes in an uncontrolled environment that we examine is able to guarantee that the elector expresses his choice alone and free from coercion.

Research to address the problems of coercion and thus to increase the respect for confidentiality were implemented in some systems by Internet voting: they offer the possibility of voting several times, the last vote being the one to be finally counted[12]. In addition, voters may vote directly at a polling station (in a controlled environment) during a few days before the official election day and cancelling their eventual vote by internet. Such attempts have the disadvantage of weakening the principle of anonymity: to enable the possibility of cancellation, votes must be stored on the server maintaining the link between the votes and the identifiers of the persons who sent them. Introducing the possibility of multiple voting eliminates coercion, a visible weakness (at least by the elector concerned) of small magnitude, but introduces a hidden vulnerability of large magnitude: the collection and analysis of the internal files to the server can reveal the identity of all voters and the choices they have made.

---

11 In France, these weaknesses had lead the service in charge with the state protection and security (Secrétariat Général de la Défense Nationale - SGDN) to advise against the use a biometrics for the security of the computer systems of the state[36].

12 Curiously, some studies suggest to give the voter the opportunity to mark a vote as final, although this possibility would destroy the benefits of multiple voting because, in case of coercion, the victim will obviously being forced to mark her forced vote as final.[35].

## IV.2 Postal voting

The weak point of the postal voting is the transmission of ballots (C2). The envelopes containing the ballots can be diverted (they are easily recognizable), or simply may not arrive in time to be used. Although postal services are expected to respect the secrecy of letters, envelopes can be opened and the votes disclosed in violation of the principle of confidentiality. There are also techniques to determine the contents of envelopes without opening them. It would be theoretically possible to use a registered delivery service and use only secure envelopes, but the enormous cost of such measures would make it unrealistic to apply them at large and there are countries where the concept of secure postal services does not exist[13].

Envelopes may be destroyed or replaced after receipt at the central polling station (B3).

These attacks on the authenticity and confidentiality principles have a medium visibility which increases with the number of letters. The difficulty of implementation also depends on the scale: it is easy to remove one or two envelopes, repeating the operation for several hundreds or thousands requires the involvement of many people which increases its visibility. In France, postal voting has been banned for political elections by the Law No. 75-1329 of 31 December 1975 [19] after many cases of proven fraud.

## IV.3 Hybrid voting

Hybrid voting has the same vulnerabilities as postal voting regarding the transmission of ballots (C2). Similarly, envelopes containing voting card can be stolen and destroyed after receipt (B3). Replacing envelopes containing voting cards is more complex than for postal voting because each voting card is unique. The magnitude of this fraud is thus limited if the voting card manufacturing is beyond the reach of the central polling station.

The counting stage (B4) is automated. Voting cards that bear the identifier and choice of each voter are scanned by a single application that manages both the updating of signing sheet and the counting of the votes. The separation between votes, identifiers and identities of voters is not clear. People with access to software that performs the counting have the possibility of disclosing the identities and choices of voters. This infringement to the vote secrecy may be done by a single person and could affect all the votes, while remaining undetected.

## IV.4 Internet voting

**Preparation and transmission of electoral material**

With Internet voting, there is no ballot paper and no voting card. A piece of information in enough to vote.

Electronic forms that contains identifiers and passwords may be copied after their generation, or at the office which prints the electoral material (B1). This operation may involve the entire electorate and not does not present great difficulties, while remaining invisible. However, to strengthen the security, information such as the date or the town of birth, are often required. The collection of this information for many people may be an insurmountable task, which limits the size of this attack.

To remove this step, and therefore the vulnerabilities that accompany it, it is possible to equip each voter with an electronic card used as an identifier. In this case one risk replaces another: the use of a single identity card[14] to perform different actions (voting, paying taxes, etc..) makes people particularly vulnerable to abusive actions of the state that might be tempted to make use of such data for undesirable purposes. This risk should not be overlooked, especially as any state that would be tempted by such practices would almost certainly not be the most willing to inform the population of the dangers[11].

---

13  Case of abroad voters.

**Between the expression of choice (E2) and the reception of votes (B3)**

These steps give rise to interactions between the voter and the server of the official vote web site.

A virus present on the voter computer can intercept the vote between its validation (E2) and its encryption (E3) and communicate it to others. It can also implement a diversion (pharming) to capture the session information entered by the voter. This information can then be used to vote at the place of the legitimate voter. These viral actions might affect a large number of votes and stay almost invisible. Their achievement does not display any particular difficulty for a motivated hacker.

Denial of service, that disrupts access to the official web site, is immediately visible.

Many people might be tempted to vote from a computer in their workplace, especially if it is a professional election, without ever realizing that companies exercise control on the use of the internet [6] and therefore might be able to spy on their employees' votes.

**Reception (B3) et counts (B4)**

During its transmission via the Internet, information about the identity of the voter and information about the choice made by the voter stay together and arrive together on the offical web site server. This point is particularly sensitive and has been the subject of numerous publications showing how to encrypt the votes in order to decode the identity of the voter independently of her choice ([13] for example). But it is still possible to reconstruct the votes from intermediate files storing the information received by the server, even if they are encrypted (having enough data and time to study facilitate this type of fraud). There is no technical measures which would make impossible to breach the secrecy of voting by a person with malicious intent and with access to the servers.

There are conventional process of fraud such as the introduction of a Trojan Horse or a Back Door. These frauds are summarized in the introduction of a few lines of program that can easily go unnoticed in the middle of programs including several thousands of lines [34]. These malpractices can be implemented by a single person. It may be a programmer, a technician responsible for maintenance and updates, or any person with a physical or logical access to the servers. The magnitude of such fraud is large.

Finally, the combination of automate updating of signing registers and the dematerialisation of ballots facilitate ballot stuffing on a large scale: at the final moments of the voting period a fraudulent program can generate many votes from voters who abstained. This risk can not be controlled by monitoring the rate of participation because it was observed that the voting sites are experiencing peak attendance in the last moments during which the vote is open. It can not be stopped by checking the voters: even if voters discover that a vote has been registered in their name when they did not vote, it will be impossible to prove.

---

14  Effective in Estonia.

# V. Assessment

## V.1 - Synthesis

None of the remote voting systems can be classified as safe. But the consequences of the vulnerabilities are heterogeneous.

### Postal voting

Postal voting is vulnerable to fraud and heavily dependant on postal services, but attacks on the fairness of elections can not go unnoticed when they are large.

### Hybrid voting

Hybrid voting delegates the counting of votes and the updating of the signing registry to automatic procedures and disallows any outside intervention in this crucial stage of an election. The procedure for counting may hide malfunctions or major frauds that keep intact the total number of counted votes but undermines its sincerity. The establishment of such a fraud would require that all received voting cards were counted again by physical persons, which is impossible if there are several thousands of ballots because of practical difficulties (you must keep the ballots sealed, find enough people to make counts, above all, be able to justify the need for such an operation) and legal (if the recount is not completed, there is no evidence to present to the election judge qualified to allow the recount).

In addition, the software may disclose to third parties how each voter voted and this violation of the secrecy of vote may be difficult to prove.

This procedure has implicit vulnerabilities of large magnitude that could stay invisible and present a small difficulty.

### Internet voting

Internet voting presents vulnerabilities of the worst kind: they are invisible, can affect a large number of votes, may be committed by a small number of people (from anywhere in the world) and do not require expensive equipment. These vulnerabilities are present at different stages of the voting process: at the voter's computer, during the delivery of votes or when the count is processed.

## V.2 - Analysis

Remote voting vulnerabilities that we examined can take place at different stages of the vote and remain undetected by officials, delegates and candidates' representatives. These stages can be corrupted without anyone knowing.

With the postal voting the areas of opacity are limited to the choice by the elector and to the transmission of letters. The automatic vote counting in hybrid voting procedures extends the areas of opacity by preventing the public counting of the votes.

Internet voting radicalises the automation process by handling dematerialised objects. The voting process is displaced from the real world to a virtual world where the observations made directly through our perceptions (sight, touch, etc..) do not apply and which is outside the reach of the majority of citizens. It is impossible to directly control the voting process and evaluate how it works correctly. It is only possible to observe processes that are supposed to reflect the activity of the voting system, but which can also give a distorted view.

The votes can be affected by events that may remain invisible: criminal acts (even easier to commit when near of the team responsible for organizing the elections) or simple malfunctions.

Lastly it appears that science is powerless against this problem. Writing a large program that hides errors can be considered as a great achievement. However, ensuring that no program hosts deliberate "errors" willfully is a far more complex task. Neither tests[15] nor the expertise of the programs are sufficient, as recalls Ken Thompson, co-designer of the UNIX system.

> « You can't trust code that you did not totally create yourself. (Especially code from companies that employ people like me.) No amount of source-level verification or scrutiny will protect you from using untrusted code. (...) As the level of program gets lower, these bugs will be harder and harder to detect. A well installed microcode bug will be almost impossible to detect. »[34]

# Conclusion

This study presented technical and democratic vulnerabilities of three ways of remote voting. It showed that the use of computerized tools lead, by nature to more complex procedures, making potentially invisible some massive attacks against authenticity or confidentiality.

It appears that the dematerialisation and the transformation of information that lie in any computerized processing put the voting process in a new world where the ordinary rules of physics no longer apply. The impossible becomes possible (thousands of votes may be altered in a moment) and the apparent banality of electronic voting systems can become a deceptive illusion. For example, in the real world the simple mixing of the anonymous envelopes breaks definitively the relationship between votes and voters. In the virtual universe, there is still no way to do this: files can be copied, information can be recovered after being erased, etc.

In France, the Commission Nationale Informatique et Liberté (CNIL) requires separate management of votes and identities but seems unaware that this separation is not likely to prohibit violations of the secrecy of the vote. Similarly, the European Commission defines transparency as the ability to verify that the voting system is functioning properly, which remains an impossible task as we have previously demonstrated. It recommends that the voter can confirm his vote and correct it[16] while this operation requires that a link between voter and vote be maintained, thus weakening the principle of anonymity. These attempts to reconcile anonymity, protection of authenticity and dematerialisation show that the transition to electronic elections conceals fundamental problems and reveals the contradictions and unexpected difficulties.

> « The clear consensus of computer-science experts around the world who have studied these issues is that Internet elections cannot be trusted, for all the reasons that I have explained: the voters and political parties cannot audit the operation of the software and hardware that serves as the real *bureau de vote*. Therefore it is not clear to me how the *assesseurs* can sign anything but a surrealist image of a true *procès-verbal*. » [1]

---

15 « Il existe en outre un théorème fondamental de la théorie de l'informatique selon lequel il ne peut y avoir de test général pour décider si un système et ses logiciels hébergent ou non un code malveillant. » (There is a foundamental information theory theorem that establishes there is no test which allows to know if a system and its softwares host, or do not host, a malware.) R. Oppliger [25]

16 « Furthermore, the elector must be able to obtain confirmation of his or her vote and, if necessary, correct it without the secrecy of the ballot being in any way violated. » [9]

Enguehard, C., Lehn R. Vulnerability analysis of three remote voting methods. XXI IPSA World Congress of Political Science, RC10 Electronic Democracy - Dilemmas of Change? Santiago, Chile, July 13, 2009.

## Bibliography

[1] APPEL (A.W.) Ceci n'est pas une urne: On the Internet vote for the Assemblée des Français de l'étranger, (juin 2006).

[2] AUER (A.), VON ARX (N.) La légitimité des procédures de vote : les défis du e-voting, faculté de droit de l'Université de Genève, Suisse, (décembre 2001).

[3] BIRDSALL (S.) The democratic divide, first monday, peer-reviewed journal on the internet, (2005).

[4] CEV. Commission on Electronic Voting, Secrecy, Accuracy and Testing of the Chosen Electronic System. first report, (December 2004).

[5] CEV. Commission on Electronic voting. Secrecy, Accuracy and Testing of the Chosen Electronic Voting System. second report, (July 2006).

[6] CNIL. La cybersurveillance des salariés. rapport de la Commission Nationale Informatique et Libertés, (2003).

[7] COLEMAN (S.) Internet voting and democratic politics in an age of crisis. in Trechsel A. (ed.) The European Union and E-Voting: Addressing The European Parliament's Internet Voting Challenge, Londres: Routledge, p.223-237, (2005)

[8] European Commission for Democracy through Law (Venise Commission). Code of Good Pratice in Electoral Matters, (juillet 2002).

[9] European Commission for Democracy through Law, (Venice Commission), Report on the compatibility of remote voting and electronic voting with the standards of the Council of Europe adopted by the Venice Commission at its 58th Plenary Session (Venice, 12-13 March 2004) , CDL-AD(2004)012.

[10] ETAT DE GENEVE. E-Voting - Cahier des charges. www.ge.ch/evoting/cahier_charges.asp

[11] DESWARTE (Y.), MALCHOR (C. A.) Current and future privacy enhancing technologies for the Internet. Ann. Télécommun., 61, n?3-4, p.399-417, (2005).

[12] DILL (D.), DOHERTY (W.) Electronic Voting Systems. Report for the National Research Council, (November 22, 2004).

[13] GOMEZ OLIVA (A.), SANCHEZ GARCIA (S.), PEREZ BELLEBONI (E.) Contributions to traditional electronic systems in order to reinforce citizen confidence. Electronic Voting 2006, 2nd International Workshop, GI-Edition, Lecture Notes in Informatics, Robert Krimmer (Ed.), p.39-49, Bregenz, Austria, (August, 2nd-4th 2006).

[14] HERRNSON (P. S.), NIEMI (R. G.), HANMER (M. J.), BEDERSON (B. B.), CONRAD (F. G.), TRAUGOTT (M.) The Importance of Usability Testing of Voting Systems. Electronic Voting Technology Workshop, Vancouver B.C., Canada, August 1, 2006.

[15] HOFF (J.) Towards a theory of Democracy for the information age. Discussion paper for the Democracy Platform UK-Nordic Meeting, (16-17 septembre 1999).

[16] HUBBERS (E.), JACOBS (B.), PIETERS (W.) RIES - Internet Voting in Action. In R. Bilof, Proceedings of the 29th Annual International Computer Software and Applications Conference, COMPSAC'05, pages 417-424. IEEE Computer Society, (July 26-28, 2005).

[17] JANVIER (R.) Lien entre modèles symboliques et computationnels pour les protocoles cryptographiques utilisant des hachages. Thèse de doctorat de l'université Joseph Fourier, Grenoble, (2006).

[18] JEFFERSON (D.R.), RUBIN (A.D.), SIMON (B.), WAGNER (D.) Analyzing Internet Voting Security. Communications of the ACM, vol.47, n?10, p.59-64, (October 2004).

[19] LOI n°75-1329 du 31 décembre 1975. codifiée sous l'article L72-1 du code électoral, (1975).

[20] MATSUMOTO (T.), MATSUMOTO (H.), K. YAMADA (K.), HOSHINO (S.) Impact of artificial "gummy" fingers on fingerprint systems, Proceedings of SPIE, Optical Security and Counterfeit Deterrence Techniques IV, vol.4677, (2002).

[21] MERCURI (R.) A Better Ballot Box?. IEEE Spectrum Online, (October 2002).

[22] MOORE (D.), PAXSON (V.), SAVAGE (S.), SHANNON (C.), STANIFORD (S.), WEAVER (N.) Inside the Slammer worm. IEEE Security and Privacy, (2003).

[23] NATIONS UNIES. Déclaration universelle des droits de l'homme, (1948).

[24] NATIONAL INSTITUTE OF STANDARDS AND TECHNOLOGY. Requiring Software Independence in VVSG 2007: STS Recommendations for the TGDC, (November 2006) Voluntary Voting System Guidelines Recommendations to the Election Assistance Commission, (August 31, 2007).

Enguehard, C., Lehn R. Vulnerability analysis of three remote voting methods. XXI IPSA World Congress of Political Science, RC10 Electronic Democracy - Dilemmas of Change? Santiago, Chile, July 13, 2009.

[25] OPPLIGER (R.) Traitement du problème de la sécurité des plates-formes pour le vote par Internet à Genève, (3 mai 2002).

[26] OSCE/ODIHR. USA 2 November 2004 Elections - OSCE/ODIHR Needs Assement Mission Report. 7-10 September 2004, Warsaw, (28 September 2004).

[27] OSCE. Election Observation Handbook, Fifth edition, ISBN 83-60190-00-3, (2005).

[28] RYAN (P.Y.A.), PEACOK (T.) Prêt à Voter: Systems Perspective, (September 20, 2005).

[29] SANDLER (D.), DERR (K.), WALLACH (D. S.) VoteBox: a tamper-evident, verifiable electronic voting system. Proceedings of the 17th USENIX Security Symposium (USENIX Security '08), (2008).

[30] SCHNEIER (B.) The Trojan Horse Race. Inside Risks 111, Communications of the ACM, vol.42, n°9, (September 1999).

[31] SERVICE CANTONAL DES VOTATIONS ET ELECTIONS. Je vote ! - élections communales, Election du Conseil municipal du 25 mars 2007. Canton de Genève, (2007).

[32] SIFAKIS (J.) cited in "In Search of Dependable Design" by Leah Hoffman. Communications of the ACM, vol.51, n°7, p.14- 16, (July 2008).

[33] SIMONS (B.) Electronic Voting Systems: the Good, the Bad, and the Stupid. ACM Queue vol.2, n°7, (October 2004).

[34] THOMPSON (K.) Reflections on Trusting Trust. Communication of the ACM, vol.27, n°8, p.761-763, (August 1984).

[35] VOLKAMER (M.), GRIMM (R.) Multiple Casts in Online Voting: Analyzing Chances. Electronic Voting 2006, 2nd International Workshop, GI-Edition, Lecture Notes in Informatics, Robert Krimmer (Ed.), p.97-106, Bregenz, Austria, (August, 2nd-4th, 2006).

[36] WOLF (P.) de l'authentification biométrique", Sécurité Informatique, n°46, p.1-6, (octobre 2003).
15